\documentclass[10pt,twocolumn,twoside]{IEEEtran}
\usepackage{algorithm}
\usepackage{algorithmic}
\usepackage{amsfonts}
\usepackage{amsmath}
\usepackage{amssymb}
\usepackage{booktabs}
\usepackage{calc}
\usepackage{cases}
\usepackage{color}
\usepackage{enumerate}
\usepackage{fixltx2e}
\usepackage{float}
\usepackage{graphicx} 
\usepackage{marvosym}
\usepackage{mathtools}
\usepackage{multicol}
\usepackage{multirow}
\usepackage[noadjust]{cite}
\usepackage{soul}
\usepackage{stfloats}
\let\sss= \scriptscriptstyle
\ifodd 1
\usepackage{soul}
\usepackage{color}
\setstcolor{red}

\newcommand{\del}[1]{\st{#1}} 

\newcommand{\com}[1]{\textbf{\color{red} (COMMENT: #1)}} 
\newcommand{\response}[1]{\textbf{\color{green} (RESPONSE: #1)}} 
\else

\newcommand{\del}[1]{}

\newcommand{\com}[1]{}
\newcommand{\comg}[1]{}
\newcommand{\response}[1]{}
\fi
\begin{document}
	
	\title{\LARGE{Spatial Modulation for Molecular Communication}}
	
	\author{\normalsize {Yu~Huang,~\IEEEmembership{}
			Miaowen~Wen,~\IEEEmembership{Senior Member,~IEEE,}
			Lie-Liang~Yang,~\IEEEmembership{Fellow,~IEEE,}\\
			Chan-Byoung~Chae,~\IEEEmembership{Senior Member,~IEEE,}
			and~Fei~Ji,~\IEEEmembership{Member,~IEEE}}
		\thanks{ 
		Y. Huang, M. Wen and F. Ji are with the School of Electronics and Information Engineering, South China University of Technology, Guangzhou 510641, China (e-mail: ee06yuhuang@mail.scut.edu.cn; eemwwen@scut.edu.cn; eefeiji@scut.edu.cn). \par
		L.-L. Yang is with the Southampton Wireless Group, School of Electronics and Computer Science, University of Southampton, Southampton SO17 1BJ, U.K. (e-mail: lly@ecs.soton.ac.uk).\par
		C.-B. Chae is with the Yonsei Institute of Convergence Technology, School of Integrated Technology, Yonsei University, South Korea (e-mail: cbchae@yonsei.ac.kr).}}
	
	\maketitle
	\vspace{-2.0cm}
\begin{abstract}
In this paper, we propose an energy-efficient spatial modulation based molecular communication (SM-MC) scheme, in which a transmitted symbol is composed of two parts, i.e., a space derived symbol and a concentration derived symbol. The space symbol is transmitted by embedding the information into the index of a single activated transmitter nanomachine. The concentration symbol is drawn according to the conventional concentration shift keying (CSK) constellation. Befitting from a single active transmitter during each symbol transmission period, SM-MC can avoid the inter-link interference problem existing in the current multiple-input multiple-output (MIMO) MC schemes, which hence enables low-complexity symbol detection and performance improvement. Specifically, in our low-complexity scheme, the space symbol is first detected by energy comparison, and then the concentration symbol is detected by the equal gain combining assisted CSK demodulation. In this paper, we analyze the symbol error rate (SER) of the SM-MC and its special case, namely the space shift keying based MC (SSK-MC), where only space symbol is transmitted and no CSK modulation is invoked. Finally, the analytical results are validated by computer simulations, and our studies demonstrate that both the SM-MC and SSK-MC are capable of achieving better SER performance than the conventional MIMO-MC and single-input single-output MC (SISO-MC) when the same symbol rate is assumed.
\end{abstract}
\ifCLASSOPTIONpeerreview
\begin{center} \bfseries EDICS Category: 3-BBND \end{center}
\fi
%
\IEEEpeerreviewmaketitle

\section{Introduction}
Chemical signalling that exploits molecule or ion for communication has been widely found in nature with diverse propagation distances. In a relatively long range, animals or insects utilize pheromone to communicate with the members of their species for mate selection, identity recognition, alarm inform, etc. \cite{Shorey.Pheromone.13,Akyildiz13}. At micro-scale environment, hormones or other chemical substances are prevalently transmitted or received in tiny organisms such as cells. This process is the so-called cell signalling that is crucial to cells' survival~\cite{Han.Signalling.17}.

Molecular communication (MC) is an emerging technique inspired from the aforementioned communication schemes in vivo, whose history dates back to 2005 \cite{Tadashi.MC.05}. MC is available at both macro-scale and micro-scale~\cite{Farsad.MCSurvey.16}. At micro-scale, MC is suitable for connection among nanomachines, whose communication distance ranges from a nanometer to a hundred nanometers. Nanomachine is one of the most remarkable progress of nanotechnology that has the potential to revolutionize many aspects of technology, and ultimately benefit our life. However, a single nanomachine can only perform very simple tasks due to its size constraint, whereas complex applications including biopsy, targeted drug delivery, environmental sensing, cell sorting, etc., require coordination among a swarm of nanomachines~\cite{Wang.Nanomachine.13}. MC is regarded as a prominent candidate for nanonetworking because of its bio-compatible and energy-efficient characteristics. It has been acknowledged as the most important communication paradigm in IEEE 1906.1 standard~\cite{Tadashi.10year.17}.

Modulation plays a significant role in MC as it determines the system's achievable performance~\cite{Suzuki.MCmodelling.17}. 
There are a few of modulation schemes for MC proposed in the literature, including the concentration shift keying (CSK)~\cite{Kuran.ModuMC.11}, molecular shift keying (MoSK)~\cite{Kuran.ModuMC.11}, isomer-based ratio shift Keying (IRSK)~\cite{Kim.Isomodu.13} and pulse position modulation (PPM)~\cite{Llatser.PPM.13}, in which messages are encoded as the concentration, type, ratio and release time of transmitted molecules, respectively. Advanced modulation schemes have also been considered in the context of the single-input single-output~(SISO) based MC (SISO-MC)~\cite{Arjmandi.MCSK.13,Kabir.DMoSK.15,Mosayebi.typesignmodu.18}. However, SISO-MC is not always suitable for the scenarios of high-speed transmission and some of the applications where reliability is indispensable. To solve these problems, some well-known techniques in the conventional wireless communications, e.g., multiple-input multiple-output (MIMO), have been redesigned for MC \cite{KC.frontierWC.12}. Specifically, MIMO based MC (MIMO-MC) is a recent trend in MC research, dated back to 2012 when it was first proposed in \cite{KC.MIMOMC.12}. It is shown that in MIMO-MC, spatial diversity enhances the bit error rate (BER) performance, while spatial multiplexing may increase transmission rate significantly~\cite{KC.MIMOMC.12}. In 2013, a micro-scale MIMO-MC system was introduced~\cite{Tadashi.MCbook.13}, where a group of sender nanomachines simultaneously transmit messages to a group of receiver nanomachines through the medium where they reside. In the same year, the first SISO-MC prototype at macro-scale was implemented \cite{Farsad.Prototype.13}, which mentioned that MIMO principle may be introduced to improve the transmission rate. In 2016, a $2 \times 2$ MIMO-MC prototype achieving spatial multiplexing was implemented, which achieves a 1.7 times higher data rate than its SISO counterpart~\cite{Chae.MCMIMOproto.16}. These research and practice demonstrate that spatial multiplexing in MIMO-MC is feasible for rate increase, although the data rate is not doubled due to the existence of interference and overhead. In~2017, MIMO-MC technique gained more attention than ever before, due to the appearance of the training-based channel estimation~\cite{Rou.CEMIMOMC.17}, spatial diversity coding techniques~\cite{Damrath.Spatialcoding.17} and the introduction of machine learning based channel modeling methods~\cite{Lee.MLMIMOMC.17}. Moreover, in \cite{WG.MCSIMOsyn.18}, synchronization was investigating in the context of the single-input multiple-output (SIMO) based MC (SIMO-MC). Now, we have no doubt that MIMO-MC constitutes a promising technique for performance improvement in MC. However, a typical challenge in MIMO-MC is the inter-link interference (ILI) in addition to the inter-symbol interference (ISI) existing in all MC systems. Due to the ILI, first, it increases the detection complexity at the receiver side. Second, the study in \cite{Chae.MCMIMOproto.16}~manifests that the BER of $2 \times 2$ MIMO-MC prototype is 2.2 times higher than that of its SISO-MC counterpart, meaning that high-reliability MC is challenging. Third, the energy consumption in MIMO-MC is significant in comparison to SISO-MC, due to more devices being simultaneously activated at both transmitter and receiver sides. This problem can be serious in micro-scale MC, since the power supply of nanomachine is limited and their computing capability is low. Consequently, the development of energy-efficient and low-complexity transmission schemes for MIMO-MC is demanding.

In this paper, we propose a spatial modulation based MC (SM-MC) for MIMO-MC implementation, which combines a space-dependent modulation with a concentration-relied modulation. By allowing only the space modulation, we also propose the special case of SM-MC, namely, the space shift keying based MC (SSK-MC). It can be shown that both SM-MC and SSK-MC are able to combat the aforementioned problems of the MIMO-MC. It is well known that SM~\cite{Mesleh.SM.08} and SSK~\cite{Jeganathan.SSK.09} in radio-based wireless communications exhibit low complexity and high energy efficiency~\cite{Wen.IMsurvey.17}. By exploiting the spatial domain for message encoding via activating an active transmit antenna in each time slot, SM and SSK are capable of eliminating the inter-antenna interference (IAI). Note that, the major distinction between SM and SSK is that SM uses both spatial and signal constellation to transmit information, while SSK only exploits the spatial constellation.

In our design for SM-MC and SSK-MC, the spatial domain is reflected by the concentration difference at the receiver side, when different transmitters emit molecules. By contrast, the signal constellation domain is contributed by the different constellation level in CSK. Hence, the SM-MC/SSK-MC receivers can collaborate to identify the activated transmitter nanomachine for detecting the space symbol, while the SM-MC receiver can decode the concentration symbol according to the rule of CSK. Furthermore, in order to enhance the reliability of detection, detector assisted by either \emph{equal gain combining} (EGC) or \emph{selection combining} (SC) is developed for detecting the concentration symbol. In this paper, we derive the SER expressions for both the SM-MC and SSK-MC, and conduct computer simulations to examine their performance as well as to validate our theoretical analysis. Our studies and results show that both the SM-MC and SSK-MC outperform the MIMO-MC in terms of the SER performance, and our analytical results are accurate enough for predicting the achieving SER performance of the SM-MC and SSK-MC systems.
 
The remainder of this paper is organized as follows. Section~II reviews the system model of MIMO-MC over diffusion channels. In Section~III, we present the principles of SM-MC and SSK-MC based on the architecture of MIMO-MC. In this section, we also derive the resulting SER of the SM-MC and SSK-MC. Section~IV evaluates the SER performance of SM-MC and SSK-MC by invoking the MIMO-MC with on-off keying (OOK) modulation and the SISO-MC with quadruple CSK (QCSK) modulation as the benchmarks. Finally, the research is concluded in Section~V.

\textit{Notation}: Boldface uppercase and lowercase letters indicate matrices and vectors, respectively. $\mathbb{R}^{n\times m}$ indicates a real number matrix with $n\times m$ dimensions. $\mathbb{E}[\cdot]$, $\left \lVert \cdot \right \rVert$, $|\cdot|$, $Q(\cdot)$ and $\text{Pr}[\cdot]$ represent expectation, Euclidean norm, absolute value, Q-function and probability of an event, respectively.

\section{Review of MIMO-MC}
We consider a MIMO-MC system in a 3-D unbounded environment with point sources and spherical receivers, which are assumed to be memoryless as \cite{Yilmaz.14,Zamiri.CDMAMC.16,Tepekule.Memoryless.15}, meaning that they have no information of the previously detected symbols. Perfect synchronization is also assumed in our MIMO-MC. In this section, we propose the channel and communication models for MIMO-MC, which constitutes the fundamentals of our subsequent SM-MC scheme and its special case of the SSK-MC scheme.
\subsection{Channel Model of MIMO-MC}
\begin{figure}[tb]
	\centering
	\includegraphics[width=3.25in]{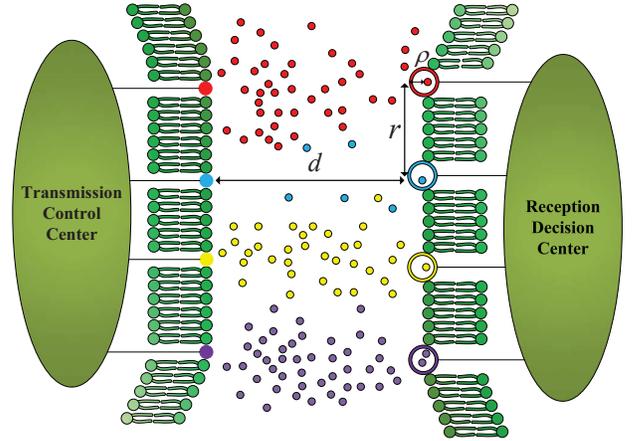}
	\caption{System model of MIMO-MC.}
	\label{fig_1}
\end{figure}

We consider an $N \times N$ diffusion based MIMO-MC system operated at micro-scale in this paper \cite{Tadashi.MCbook.13}, where $N$ transmit and receive nanomachines are attached to the cell membrane at both transmitter and receiver sides, respectively. Furthermore, we assume that there are a transmission control center and a reception decision center, respectively, located at the centers of the transmitter and receiver cells. The function of transmission control center is to coordinate the transmitter nanomachines to emit molecular pulses according to the information to be transmitted, while the reception decision center connecting all the receiver nanomachines decodes the information based on the received signals from $N$ receiver nanomachines. To clearly demonstrate the communication system model of the MIMO-MC, we exemplify a $4 \times 4$ MIMO-MC system with $4$ pairs of transceivers as shown in Fig. 1, where the lipid bilayer of cell membrane are shown in green color. As shown in Fig. 1, the pair of transmitter and receiver of a link and the molecules transmitted between them are marked using the same color for the sake of easy distinction. Under the assumption that perfect alignment is achieved in the MIMO-MC architecture, the spacing between adjacent transmitters or adjacent receivers is equally set as $r$, and the distance between a pair transmitter and receiver is expressed as $d$. Thus, the distance from the $j$-th transmitter to the $i$-th receiver is given by
\begin{align}
d_{ji} =
\begin{cases}
\hspace{1cm}d,  & \text{for $i=j$}, \\
\sqrt{d^{2}+|j-i|^{2}r^{2}}, & \text{for $i\neq j$.}
\end{cases}
\end{align}

In this paper, we assume $M$-ary CSK modulation and denote the concentration of molecules in the $i$-th receiver at time $t$ in response to the $j$-th transmitter as $c_{m,ij}(t)$ when a pulse of $S_{m}$ molecules with $m\in\{0,1,\ldots, M-1\}$ is emitted at $t=0$. Consequently, the concentration at receiver can be formulated according to Fick's second law of diffusion as \cite{Dk.receiverMC.13}
\begin{align}
c_{m,ij}(t)=S_{m}\frac{1}{{(4\pi Dt)}^\frac{3}{2}}\exp \left(-\frac{d_{ji}^2}{4Dt}\right),
\begin{array}{c}
\hspace{-0.3cm}j\in\{1,2,\ldots, N\},\\
\hspace{-0.3cm}i\in\{1,2,\ldots, N\},\\
\hspace{-0.4cm}m\in\{0,1,\ldots, M-1\},\\
\end{array}
\end{align}
where $D$ is the diffusion coefficient of the propagation medium that is assumed to be homogeneous in this paper. We assume that the spherical receiver nanomachine has a volume of $V_{\sss \text{RX}}=\frac{4}{3}\pi{\rho}^{3}$ with $\rho$ being the radius of the receiver. Molecular concentration is assumed to be uniform inside a passive receiver when $d\gg\rho$. Based on these assumptions, the expected number of molecules inside the $i$-th receiver at time $t$ can be formulated by \cite{MM.studyofCEMC.14,ANoel.Passivereceiver.16}
\begin{align}
N_{m,ij}(t)=V_{\sss\rm{RX}}c_{m,ij}(t)=V_{\sss\rm{RX}}S_{m}h_{ij}(t),&\text{\quad\quad for $t>0,$}
\end{align}
where $h_{ij}(t)$ indicates the probability that an information molecule released at $t=0$ is sensed by the passive receiver at time~$t$ \cite{GeC.adadetec.18,GD.AmpliDetect.17}, which is given by
\begin{align}
h_{ij}(t)=\frac{1}{{(4\pi Dt)}^\frac{3}{2}}\exp\left(-\frac{d_{ji}^2}{4Dt}\right).
\end{align}
In MC, $h_{ij}(t)$ represents the channel impulse response (CIR) between the $j$-th transmitter and the $i$-th receiver. Therefore, the concentration vector $\mathbf{c}_{m,j}(t)$$\in \mathbb{R}^{N \times 1}$ collecting the expected concentration of all receivers in response to the $j$-th transmitter can be written as
\begin{align}
\mathbf{c}_{m,j}(t)&\triangleq\big[c_{m,1j}(t), \ldots, c_{m,ij}(t), \ldots, c_{m,Nj}(t)\big]^{T}\nonumber\\
&=S_{m}\big[h_{1j}(t), \ldots, h_{ij}(t), \ldots, h_{Nj}(t)\big]^{T}\nonumber\\
&=S_{m}\mathbf{h}_{j}(t),
\end{align}
where $\mathbf{h}_{j}(t)$ is the CIR vector from the $j$-th transmitter to all the $N$ receivers at time $t$. Let us define $\mathbf{H}(t)$$\in \mathbb{R}^{N \times N}$ as the channel matrix of an $N \times N$ MIMO-MC system at time $t$, which can be represented as
\begin{align}
\mathbf{H}(t)&=\Big[\mathbf{h}_{1}(t); \ldots ;\mathbf{h}_{j}(t); \ldots ; \mathbf{h}_{N}(t)\Big],
\end{align}
whose entries are given by (4). Based on our assumptions, we can know that the diagonal elements of $\mathbf{H}(t)$, such as $h_{jj}(t)$, have the same value, given by (4) associated with setting $i=j$ and $d_{jj}=d$. By contrast, a non-diagonal element $h_{ij}(t)$ gives the probability of the ILI from the transmitter $j$ to the receiver~$i$.

In this paper, we consider the amplitude detection \cite{Llatser.PPM.13}. To achieve this, we assume that all the receivers sense the concentration at a certain time, e.g., at the time when the concentration at a receiver generated by its paired transmitter reaches the peak value, which can be obtained by solving the partial derivative equation $\frac{\partial{c_{m,jj}(t)}}{\partial{t}}=0$. Specifically, when an impulse of molecules is emitted by a transmitter at $t=0$, the peak concentration presenting at its paired receiver can be found to be at the time of
\begin{align}
t_{p}=\frac{d^{2}}{6D}.
\end{align}
Explicitly, the peak time is irrelevant to $S_{m}$. In this paper, without further explanation, it is assumed that each receiver of the MIMO-MC samples for concentration after a time interval $t_{p}$ seconds from the emission of the chemical impulse by its paired transmitter. Therefore, upon substituting (1) and (7) into (2), we can derive the maximum concentration of a receiver in response to its paired transmitter as
\begin{align}
c_{m,jj}(t_{p})=S_{m}\bigg(\frac{3}{2\pi ed^{2}}\bigg)^{\frac{3}{2}},
\end{align}
when an impulse of $S_{m}$ molecules is released. Similarly, $c_{m,ij}(t_{p})$ can be derived, given by
\begin{align}
c_{m,ij}(t_{p})=S_{m}\bigg(\frac{3}{2\pi d^{2}}\bigg)^{\frac{3}{2}}\exp \left(-\frac{3d_{ji}^2}{2d^2}\right).
\end{align}

\subsection{Communication Model of MIMO-MC}
\begin{figure}[!t]
	\centering
	\includegraphics[width=3.25in]{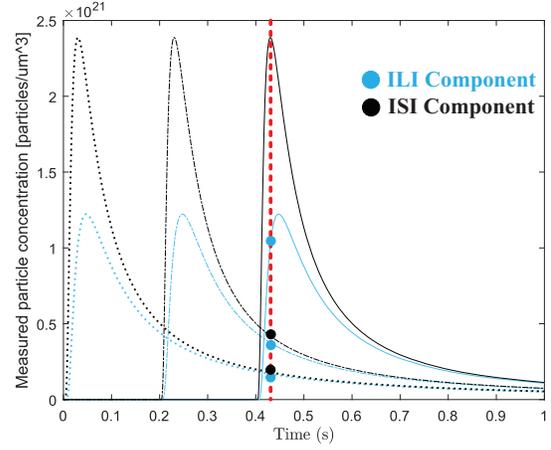}
	\caption{Concentration expected at one receiver of a $2\times2$ MIMO-MC affected by its ISI and the ILI from one other link, when $\text{SNR}=10\ \rm{dB}$, $D=2.2\times10^{-9}$$m^{2}/s$, $d=20um$, $r=15um$ and $T_{s}=0.2s$.}
	\label{fig_2}
\end{figure}

Binary CSK (BCSK) is the simplest CSK with $M=2$, which emits a chemical pulse containing $S_{1}$ molecules towards its paired receiver for transmitting bit ``1'', or a pulse of $S_{0}$ molecules for transmitting bit ``0''. As a special case of BCSK, the OOK modulation, which is prevalently adopted in the existing MIMO-MC \cite{KC.MIMOMC.12,Chae.MCMIMOproto.16,Lee.MLMIMOMC.17,Damrath.Spatialcoding.17}, keeps silent without any molecule emission for bit ``0'', i.e., $S_{0}=0$. For the general CSK, it has been revealed in literature that the CSK with $M\geq4$ is usually unable to attain satisfactory error performance in SISO-MC \cite{Suzuki.MCmodelling.17,Kabir.DMoSK.15}. Note that in Fig. 1, we considered OOK as the modulation scheme for MIMO-MC, where the red, yellow, and purple colored links represent transmitting $S_{1}$ molecules to their corresponding receivers, while the blue colored one keeps silent without any emission of molecules. Furthermore, in Fig. 1, there are still a few of blue-colored molecules, which are the residual molecules of the previous transmissions. 

Let us denote the transmit signal vector at the sampling time $t$ as
\begin{align}
\mathbf{x}(t)=\Big[x_{1}(t),\ldots,x_{i}(t),\ldots, x_{N}(t)\Big]^{T},
\end{align}
where $x_{i}(t)$$\in \{S_{0},S_{1}\}$ denotes the number of molecules emitted by the $i$-th transmitter. Then, based on (6) and (9), the concentration vector corrupted by noise sensed at time $t$ can be expressed as
\begin{align}
\mathbf{y}_{\sss \text{MIMO}}(t)&=\Big[{y}_{1,\sss \text{ MIMO}}(t),\ldots,{y}_{i,\sss \text{ MIMO}}(t),\ldots, {y}_{N,\sss \text{ MIMO}}(t)\Big]^{T}\nonumber\\
&=\sum_{l=0}^{L}\mathbf{H}(t+lT_{s})\mathbf{x}(t-lT_{s})+{\mathbf{n}_{\sss \text{MIMO}}}(t),
\end{align}
where $y_{i,\sss \text{ MIMO}}(t)$ represents the concentration sensed by the $i$-th receiver at time $t$, $T_{s}$ is the symbol duration. We assume that the ISI and ILI last for $L$ and $L+1$ symbol durations, respectively, where ISI is generated by the paired transmitter, while the ILI is the interference resulted from the other links. Practically, there is also noise in MC, which is given by the form of the noise vector ${\mathbf{n}_{\sss \text{MIMO}}}(t)$$\in \mathbb{R}^{N \times 1}$, having
\begin{align}
\mathbf{n}_{\sss \text{MIMO}}(t)=[n_{1,\sss \text{ MIMO}}(t),\ldots,n_{i,\sss \text{ MIMO}}(t),\ldots, n_{N,\sss \text{ MIMO}}(t)]^{T},
\end{align}
where $n_{i,\sss \text{ MIMO}}(t)$ is a signal dependent noise sensed by the $i$-th receiver. Similar to \cite{Damrath.Spatialcoding.17}, $y_{i,\sss \text{ MIMO}}(t)$ is the concentration sensed by the $i$-th receiver at time $t$:
\begin{align}
y_{i,\sss \text{ MIMO}}(t)= \sum_{j=1}^{N}\sum_{l=0}^{L}{h}_{ij}(t+lT_{s}){x}_{j}(t-lT_{s})+n_{i,\sss \text{ MIMO}}(t).
\end{align}

The impact of ISI and ILI on a desired signal is illustrated in Fig. 2 for a $(2\times2)$ MIMO-MC system, where the ILI is not only from the previous molecular symbols, but also from the current one released by the unpaired transmitter. However, it is worth noting that the current ILI plays a significant role in confusing the detection of a desired molecular signal, since the current ILI, as shown in Fig. 2, generates the strongest interference. The existing studies on SISO-MC often ignore the effect of ISI or only take one previous symbol into account \cite{AN.Enzyme.14}, due to the fact that this previous symbol generates the highest ISI. However, when MIMO-MC is considered, as shown in Fig.~2, the effect of the current ILI may significantly surpass that of the ISI. In other words, the interference in MIMO-MC system may be dominated by the current ILI generated by the other unpaired transmitters, and the ISI generated by one previous symbol transmitted by the paired transmitter. Therefore, if we express (13) as
\begin{align}
y_{i,\sss \text{ MIMO}}(t)=
\underbrace{{h}_{ii}(t){x}_{i}(t)}_{\text{desired signal}}+\underbrace{I_{i,\sss \text{ MIMO}}(t)}_{\text{sum of interference}}+\underbrace{n_{i,\sss \text{ MIMO}}(t)}_{\text{noise}},
\end{align}
the noise component can be assumed to follow the Gaussian distribution, depending on the current symbol at time $t$ as \cite{MengLS.receiverdesign.14}:
\begin{align}
n_{i,\sss \text{ MIMO}}(t)\sim \mathcal{N}\big(\mu_{ni,\sss \text{ MIMO}}(t),\sigma^{2}_{ni,\sss \text{ SM}}(t)\big),
\end{align}
with
\begin{align}
\mu_{ni,\sss \text{ MIMO}}(t)=0,\quad \sigma^{2}_{ni,\sss \text{ SM}}(t)=\frac{{h}_{ii}(t){x}_{i}(t)+I_{i,\sss \text{ MIMO}}(t)}{V_{\sss \text{RX}}}.\nonumber
\end{align}

The interference component in (14) can be approximated as
\begin{align}
I_{i,\sss \text{ MIMO}}(t)\approx\underbrace{\sum\nolimits_{j\neq i}{h}_{ij}(t){x}_{j}(t)}_{\text{current ILI}}+\underbrace{{h}_{ii}(t+T_{s}){x}_{i}(t-T_{s})}_{\text{last ISI}},
\end{align}
when considering that the ILI is only from the current unpaired transmitter, while the ISI is only from one previous symbol sent by the paired transmitter.
\section{Spatial modulation based molecular communication}

From the above description, we can know that MIMO-MC experiences both ILI and ISI, which may significantly degrade the detection performance. In order to combat these problems, in this section, we propose the spatial modulation based MC (SM-MC) as one of the implementation of the MIMO-MC. In our proposed scheme, both spatial and concentration domains are exploited for conveying information simultaneously, but only one type of molecules is used for transmission.
\subsection{Transmitter of SM-MC}
\begin{figure}[t]
	\centering
	\includegraphics[width=3.2in]{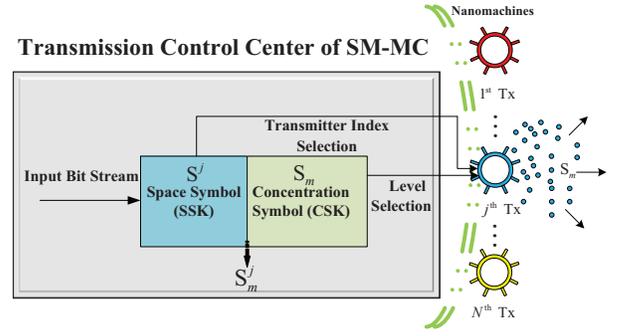}
	\caption{Transmitter diagram of SM-MC.}
	\label{fig_3}
\end{figure}

The ideology of SM-MC is inspired by the SSK technique having been widely studied in MIMO communications. Typically, in SSK modulation, only one transmit antenna is activated during each symbol period. At the receiver, the index of the activated transmit antenna can be detected, when the channel state information is available. The SSK modulation can be implemented in conjunction with a conventional amplitude-phase modulation, forming the SM \cite{LLY.PSM.11}. Similarly, the SM-MC proposed in this paper is the combination of a SSK modulation and a CSK modulation. Specifically, given a symbol transmitted, one of the transmitters releases a pulse of molecules, with the number of molecules determined also by the data symbol being transmitted. In detail, the transmit schematic diagram for our SM-MC system is depicted in Fig.~3. Hence, when the $j$-th space symbol is transmitted, with $j\in\{1,2,\ldots, N\}$, the transmit signal vector has a form of
\begin{align}
\mathbf{x}(t)=\Big[0,0,\ldots,\underbrace{S_{m}}_{\text{j-th transmitter}},\ldots, 0,0\Big]^{T},
\end{align}
where only the $j$-th transmitter is activated to emit $S_{m}>0$ molecules, when the $m$-th concentration symbol $S_{m}$ is selected. The space symbol is denoted as $S^{j}$, when the $j$-th transmitter is activated. 
We assume that the space symbols $S^{j}$ and the concentration symbols $S_{m}$ are independent of each other, solely depended on the input data stream. Then, we have
\begin{align}
\text{Pr}[S_{m}]=\frac{1}{M},\quad m\in\{0,1,\ldots, M-1\},
\end{align}
and
\begin{align}
\text{Pr}[S^{j}]=\frac{1}{N},\quad j\in\{1,2,\ldots, N\}.
\end{align}

Furthermore, if we express the SM symbols representing the combinations of the space and concentration symbols as $S^{j}_{m}$, we have
\begin{align}
\text{Pr}[S^{j}_{m}]=\frac{1}{MN},
\begin{array}{c}
j\in\{1,2,\ldots, N\},\\
m\in\{0,1,\ldots, M-1\}.\\
\end{array}
\end{align}

Therefore, the data rate of the SM-MC measured in bits per molecular symbol is given as
\begin{align}
R_{\sss \text{SM}}=\log_{2}N+\log_{2}M,
\end{align}
where both the values of $N$ and $M$ are assumed to be an integer power of $2$.

Based on (17), the concentration signal observed at the $i$-th receiver in SM-MC systems is similar to (14), expressed as
\begin{align}
y_{i,\sss \text{ SM}}(t)=S_{m}{h}_{ij}(t)+{I}_{i, \sss \text{ SM}}(t)+n_{i,\sss \text{ SM}}(t),
\end{align}
where both $S_{m}{h}_{ij}(t)$ and $n_{i,\sss \text{ SM}}(t)$ are dependent on the current molecular symbol being received, representing the expected number of molecules received and the noise component, respectively, at time $t$, when the $j$-th transmitter is activated to emit a chemical impulse with $S_{m}$ molecules. However, unlike the case in MIMO-MC \cite{Chae.MCMIMOproto.16}, the ${I}_{i, \sss \text{ SM}}(t)$ in (22) consists of only the ISI component that is resulted from a previous molecular symbol emitted by the $i$-th transmitter, since we assume that ILI only occurs with the current transmission, while ISI is only experienced from one previous emission by a paired transmitter. Let us denote the previous molecular symbol as $S^{\bar{j}}_{\bar{m}}$, then we have
\begin{align}
{I}_{i, \sss \text{ SM}}(t)=
\begin{cases}
\ 0, & \text{for $i\neq j$}, \\
\ S_{\bar{m}}{h}_{ii}(t+T_{s}),& \text{for $i= \bar{j}$.}
\end{cases}
\end{align}

Similar to (15), the distribution of the noise component in (22) can be expressed as
\begin{align}
n_{i,\sss \text{ SM}}(t)\sim \mathcal{N} \Big(\mu_{ni,\sss \text{ SM}}(t),\sigma^{2}_{ni,\sss \text{ SM}}(t)\Big),
\end{align}
associated with
\begin{align}
\mu_{ni,\sss \text{ SM}}(t)=0,\quad \sigma^{2}_{ni,\sss \text{ SM}}(t)=\frac{S_{m}{h}_{ij}(t)+{I}_{i, \sss \text{ SM}}(t)}{V_{\sss \text{RX}}}.\nonumber
\end{align}

Furthermore, based on (22), (23) and (24), the distribution of $y_{i,\sss \text{ SM}}(t)$ can be found, and expressed as
\begin{align}
y_{i,\sss \text{ SM}}(t)\hspace{-0.1cm}\sim\hspace{-0.05cm}\mathcal{N}\Big(\hspace{-0.1cm}S_{m}{h}_{ij}(t)+{I}_{i, \sss \text{ SM}}(t),\sigma^{2}_{ni,\sss \text{ SM}}(t)\hspace{-0.05cm}\Big).
\end{align}

\begin{table}[t]
	\caption{Bits per pulse for different modulation schemes in MC}  
	\begin{tabular*}{9cm}{lclc}  
		\toprule 
		Scheme & Bits per pulse&Scheme & Bits per pulse   \\  
		\toprule  
		BMoSK& 1&$N\times N$ MIMO (OOK)  & 2\\
		BCSK & 1&$2\times2$ SM (BCSK)& 2 \\
		BSSK & 1&$8$SSK & 3\\
		OOK  & 2&$2\times2$ SM (QCSK)& 3\\
		QCSK & 2&$4\times4$ SM (BCSK)& 3\\
		QSSK & 2&$16$SSK & 4\\ 
		\bottomrule  
	\end{tabular*}  
\end{table}  
From the above analysis, we can readily realize that our proposed SM-MC transmission scheme is capable of mitigating the ILI experienced by the general MIMO-MC \cite{Chae.MCMIMOproto.16}. Furthermore, the SM-MC scheme is more energy-efficient than the MIMO-MC. In MIMO-MC, when fixing a modulation scheme, increasing $N$ may transmit more bits per symbol but at the expense of higher energy consumption. By contrast, SM-MC and its special scheme of SSK-MC can obtain a logarithmic increase in bits per pulse transmission, which is contributed by the space modulation, with the increase of $N$, but without demanding any extra transmission energy. In order to highlight the energy efficiency achieved by the SM-MC and SSK-MC schemes, we compare the different modulation techniques of MC in Table~I. Note that in Table~I, we assume that the input data are independently uniformly distributed binary data. Hence, for example, when the OOK is used, transmitting $M$ bits requires about $M/2$ pulses. Therefore, we have the average bits per pulse equaling 2. As another example, when the $N\times N$ MIMO-MC using OOK is considered, it also has the average bits per pulse equaling 2.

\subsection{Signal Detection in SM-MC Systems}
In theory, the maximum-likelihood (ML) detection can be implemented by the SM-MC to jointly decode the index of the activated transmitter and the concentration symbol. When the memoryless receivers are taken into account, the ML detection based on (22) can be formulated as
\begin{align}
<\hat{j},\hat{m}>=\mathop{\arg\min}_{j\in\{1,2,\dots,N\},m\in\{0,1,\dots,M-1\}}\left \lVert\mathbf{y}_{\sss \text{SM}}(t)-S_{m}\mathbf{h}_{j}(t)\right \rVert^{2},
\end{align}
where $\hat{j}$ and $\hat{m}$ are the estimated indices of the space and concentration symbols, respectively. Note that $S_{m}$ in SM-MC is always greater than zero, i.e., $S_{m}>0$, otherwise the space symbol is unable to be detected. However, (26) imposes a search complexity of $\mathcal{O}(NM)$, when $NM$ is relatively large, hence it is not practical to deploy in MC systems, due to the constraint of the size and computing power of MC receivers. To this end, we suggest a low-complexity successive detection based on a scheme that has been commonly used in SM for wireless communications \cite{JJ.optdetectSM.08,RZ.GPSM.13}. In this scheme, the CSK symbol is detected after the detection of the space symbol. To be more specific, we detect the index of the activated transmitter via the concentration comparison of the $N$ receivers based on the fact that the receiver paired with the activated transmitter is most likely to have the maximum concentration at the sampling time, as it is located with the minimum distance from its paired transmitter. Accordingly, the detection of the space symbol can be formulated as
\begin{align}
\hat{j}= \mathop{\arg\max}_{j\in\{1,2,\dots,N\}}\;{y_{j,\sss \text{ SM}}(t)}.
\end{align}
After the detection of the space symbol, the $\hat{j}$-th receiver's concentration can be used to detect the concentration symbol, which can be described as
\begin{align}
\hat{m}=\mathop{\arg\min}_{m\in\{0,1,\dots,M-1\}}|{y}_{\hat{j},\sss \text{ SM}}(t)-S_{m}h_{{\hat{j}\hat{j}}}(t)|^{2},
\end{align}

From (27) and (28), we can readily know that the complexity of our proposed detector is $\mathcal{O}(N+M)$. Provided that $N\geqslant2$, $M\geqslant2$, the complexity of the proposed detection scheme is lower than that of the ML detection of (26). It is worth noting that, once there is a transmitter activated, all the $N$ receivers can collect the molecules released. Hence, all the receivers can collaborate to make more reliable detection of the CSK symbol at the second stage. We refer to this detector as the EGC, described as
\begin{align}
\hat{m}_{\sss \text{EGC}}&=\mathop{\arg\min}_{m\in\{0,1,\dots,M-1\}}\left \lVert\mathbf{y}_{\sss \text{SM}}(t)-S_{m}\mathbf{h}_{\hat{j}}(t)\right \lVert^{2}.
\end{align}
It can be shown that the complexity of (29) combined with (27) is still $\mathcal{O}(N+M)$, although the total number of computation of (29) is significantly higher than that of (28). However, the performance of the EGC-assisted detection can significantly outperform that of the detector of (28), which will be validated by the simulation results in Section IV.
\subsection{Error Performance Analysis of SM-MC}
In this subsection, we focus on the SER analysis of the successive detection, given by (27) and (29). Since the detection is dependent on the current molecular symbol $S^{j}_{m}$ and the previous one $S^{\bar{j}}_{\bar{m}}$, let us denote the correct detection probability of both space and concentration symbols as $P_{c,\sss \text{ SSK}}(S^{j}_{m}|S^{\bar{j}}_{\bar{m}})$ and $P_{c,\sss \text{ CSK}}(S^{j}_{m}|S^{\bar{j}}_{\bar{m}})$, respectively. Remembering that the $S^{j}_{m}$ and the $S^{\bar{j}}_{\bar{m}}$ are mutually independent, we can express the correct detection probability of space symbol as
\begin{align}
P_{c,\sss \text{ SSK}}(S^{j}_{m}|S^{\bar{j}}_{\bar{m}})=\prod_{i\neq j}Q\left(-\frac{\mu_{ji|S^{\bar{j}}_{\bar{m}}}}{\sigma_{ji|S^{\bar{j}}_{\bar{m}}}}\right),
\end{align}
where $Q(\cdot)$ is the Q-function. When given $i\neq j$, $\mu_{ji|S^{\bar{j}}_{\bar{m}}}$ and $\sigma^{2}_{ji|S^{\bar{j}}_{\bar{m}}}$ are shown as
\begin{align}
\mu_{ji|S^{\bar{j}}_{\bar{m}}}\hspace{-0.15cm}=\hspace{-0.15cm}
\begin{cases}
\vspace{0.15cm} 
\
\hspace{-0.225cm}S_{m}\hspace{-0.075cm}\big(h_{jj}\hspace{-0.05cm}(t)\hspace{-0.1cm}-\hspace{-0.1cm}h_{ij}\hspace{-0.05cm}(t)\hspace{-0.06cm}\big)\hspace{-0.1cm}-\hspace{-0.1cm}S_{\bar{m}}h_{ii}\hspace{-0.05cm}(t\hspace{-0.1cm}+\hspace{-0.1cm}T_{s}\hspace{-0.05cm}),\hspace{-0.3cm}&\text{for $i= \bar{j}$ and $j\neq \bar{j}$,}\\
\
\vspace{0.15cm}  \hspace{-0.225cm}S_{m}\hspace{-0.075cm}\big(h_{jj}\hspace{-0.05cm}(t)\hspace{-0.1cm}-\hspace{-0.1cm}h_{ij}\hspace{-0.05cm}(t)\hspace{-0.06cm}\big)\hspace{-0.1cm}+\hspace{-0.1cm}S_{\bar{m}}h_{jj}\hspace{-0.05cm}(t\hspace{-0.1cm}+\hspace{-0.1cm}T_{s}\hspace{-0.05cm}),\hspace{-0.3cm}&\text{for $i\neq \bar{j}$ and $j= \bar{j}$,}\\
\ \hspace{-0.225cm}S_{m}\hspace{-0.075cm}\big(h_{jj}\hspace{-0.05cm}(t)\hspace{-0.1cm}-\hspace{-0.1cm}h_{ij}\hspace{-0.05cm}(t)\hspace{-0.06cm}\big), \hspace{-0.3cm}&\text{for $i\neq \bar{j}$ and $j\neq \bar{j}$.}\nonumber
\end{cases}
\end{align}
\begin{align}
\sigma^{2}_{ji|S^{\bar{j}}_{\bar{m}}}\hspace{-0.15cm}=\hspace{-0.15cm}
\begin{cases}
\vspace{0.15cm}
\ \hspace{-0.25cm}\frac{S_{m}\big(h_{jj}(t)+h_{ij}(t)\big)+S_{\bar{m}}h_{ii}(t+T_{s})}{V_{\sss\text{RX}}},&\text{for $i= \bar{j}$ and $j\neq \bar{j}$,}\\
\vspace{0.15cm}
\ \hspace{-0.25cm}\frac{S_{m}\big(h_{jj}(t)+h_{ij}(t)\big)+S_{\bar{m}}h_{jj}(t+T_{s})}{V_{\sss\text{RX}}},&\text{for $i\neq \bar{j}$ and $j= \bar{j}$,}\\
\ \hspace{-0.25cm}\frac{S_{m}\big(h_{jj}(t)+h_{ij}(t)\big)}{V_{\sss\text{RX}}},&\text{for $i\neq \bar{j}$ and $j\neq \bar{j}$.}\nonumber
\end{cases}
\end{align}

\begin{IEEEproof}
	See Appendix A.
\end{IEEEproof} 

Specifically, when only the space symbol $S^{j}$ is transmitted, the SM-MC with $M=1$ degenerates to the SSK-MC and the detection in (27) is non-coherent. The detection performance of the SSK-MC is solely described by the expectation of $P_{c,\sss \text{ SSK}}(S^{j}|S^{\bar{j}})$, where $S^{j}$ and $S^{\bar{j}}$ are also mutually independent. Note that in SSK-MC, both $S^{j}$ and $S^{\bar{j}}$ are integers. Correspondingly, considering that the transmitted symbols are uniform distribution, we have
\begin{align}
P_{c,\sss \text{ SSK}}&=\mathbb{E}\Big[\mathbb{E}\big[P_{c,\sss \text{ SSK}}(S^{j}|S^{\bar{j}})\big]\Big]\nonumber\\
&=\sum_{j=1}^{N}\sum_{\bar{j}=1}^{N}P_{c,\sss \text{ SSK}}(S^{j}|S^{\bar{j}})\text{Pr}[S^{\bar{j}}]\text{Pr}[S^{j}]\nonumber\\
&=\frac{1}{N^{2}}\sum_{j=1}^{N}\sum_{\bar{j}=1}^{N}\prod_{i\neq j}Q\left(\frac{-\mu_{ji|S^{\bar{j}}}}{\sigma_{ji|S^{\bar{j}}}}\right).
\end{align}
Note that in (31), the subscript $m$ is omitted, since there is no concentration symbol transmitted in SSK-MC. Furthermore, the error probability can be easily derived as
\begin{align}
P_{e,\sss \text{ SSK}}=1-P_{c,\sss \text{ SSK}}=1-\frac{1}{N^{2}}\sum_{j=1}^{N}\sum_{\bar{j}=1}^{N}\prod_{i\neq j}Q\left(\frac{-\mu_{ji|S^{\bar{j}}}}{\sigma_{ji|S^{\bar{j}}}}\right).
\end{align}

Having considered the space symbol, the error rate of the CSK detection can be analyzed based on the estimate $\hat{j}$. Here, we analyze the more general EGC-assisted detector of (29), which utilizes all receivers' observations to detect the CSK symbol. Given that a pulse of $S_{m}$ molecules is transmitted by the $j$-th transmitter and that the previously emitted molecular symbol is $S^{\bar{j}}_{\bar{m}}$, the probability of erroneous detection of the CSK symbol can be upper bounded by
\begin{align}
P_{e,\sss \text{ CSK}}(S^{j}_{m}|S^{\bar{j}}_{\bar{m}})&\leqslant \sum_{m\neq n}\text{Pr}[S_{m}\rightarrow S_{n}|S^{\bar{j}}_{\bar{m}}],
\end{align}
where $\text{Pr}[S_{m}\rightarrow S_{n}|S^{\bar{j}}_{\bar{m}}]$ is also dependent on the detection of the space symbol. Hence, it can be expressed as
\begin{align}
\text{Pr}[S_{m}\rightarrow S_{n}|S^{\bar{j}}_{\bar{m}}]=&\sum_{\hat{j}\neq j}\text{Pr}_{\sss \text{ SSK}}[\hat{j}|S^{\bar{j}}_{\bar{m}}]\ \text{Pr}[S_{m}\rightarrow S_{n}|S^{\bar{j}}_{\bar{m}},\hat{j}]\nonumber\\
&+\text{Pr}_{\sss \text{ SSK}}[j|S^{\bar{j}}_{\bar{m}}]\ \text{Pr}[S_{m}\rightarrow S_{n}|S^{\bar{j}}_{\bar{m}},j],
\end{align}
where $\text{Pr}_{\sss \text{ SSK}}[\hat{j}|S^{\bar{j}}_{\bar{m}}]$ and $\text{Pr}_{\sss \text{ SSK}}[j|S^{\bar{j}}_{\bar{m}}]$ have been derived in (43) of Appendix~A for the detection of the SSK symbol. Hence, below we only focus on the unknown components, i.e., $\text{Pr}[S_{m}\rightarrow S_{n}|S^{\bar{j}}_{\bar{m}},\hat{j}]$ and $\text{Pr}[S_{m}\rightarrow S_{n}|S^{\bar{j}}_{\bar{m}},j]$, in (34), which are given by (35) and (36), respectively, shown on the top of the next page. In these formulas, the $\sigma^{2}_{ni,\sss \text{ SM}}(t)$ component has been defined in (24).
\begin{figure*}[t]
	\begin{align}
	&\text{Pr}[S_{m}\rightarrow S_{n}|S^{\bar{j}}_{\bar{m}},\hat{j}]=\text{Pr}\bigg[\sum_{i=1}^{N}|{y}_{i,\sss \text{ SM}}(t)-S_{m}{h}_{i\hat{j}}(t)|^{2}>\sum_{i=1}^{N}|{y}_{i,\sss \text{ SM}}(t)-S_{n}{h}_{i\hat{j}}(t)|^{2}\Big|S^{\bar{j}}_{\bar{m}}\bigg]\nonumber\\
	&=
	\begin{cases}
	\ 1-Q\left(\frac{-S_{\bar{m}}{h}_{\bar{j}\hat{j}}(t){h}_{\bar{j}j}(t+T_{s})-S_{m}\sum_{i=1}^{N}{h}_{i\hat{j}}(t){h}_{ij}(t)+\frac{(S_{m}+S_{n})}{2}\sum_{i=1}^{N}{h}^{2}_{i\hat{j}}(t)}{\sqrt{\sum_{i=1}^{N}{h}^{2}_{i\hat{j}}(t)\sigma^{2}_{ni,\sss \text{ SM}}(t)}}\right),&\text{for $S_{m}>S_{n}$}, \\
	\ Q\left(\frac{-S_{\bar{m}}{h}_{\bar{j}\hat{j}}(t){h}_{\bar{j}j}(t+T_{s})-S_{m}\sum_{i=1}^{N}{h}_{i\hat{j}}(t){h}_{ij}(t)+\frac{(S_{m}+S_{n})}{2}\sum_{i=1}^{N}{h}^{2}_{i\hat{j}}(t)}{\sqrt{\sum_{i=1}^{N}{h}^{2}_{i\hat{j}}(t)\sigma^{2}_{ni,\sss \text{ SM}}(t)}}\right),&\text{for $S_{m}<S_{n}$}.
	\end{cases}
	\end{align}
	\begin{align} 
	&\text{Pr}[S_{m}\rightarrow S_{n}|S^{\bar{j}}_{\bar{m}},j]
	=\text{Pr}\bigg[\sum_{i=1}^{N}|{y}_{i,\sss \text{ SM}}(t)-S_{m}{h}_{ij}(t)|^{2}>\sum_{i=1}^{N}|{y}_{i,\sss \text{ SM}}(t)-S_{n}{h}_{ij}(t)|^{2}\Big|S^{\bar{j}}_{\bar{m}}\bigg]\nonumber\\
	&=
	\begin{cases}
	\ 1-Q\left(\frac{-S_{\bar{m}}{h}_{\bar{j}\hat{j}}(t){h}_{\bar{j}j}(t+T_{s})+\frac{S_{n}-S_{m}}{2}\sum_{i=1}^{N}{h}^{2}_{ij}(t)}{\sqrt{\sum_{i=1}^{N}{h}^{2}_{ij}(t)\sigma^{2}_{ni,\sss \text{ SM}}(t)}}\right),  & \text{for $S_{m}>S_{n}$}, \\
	\ Q\left(\frac{-S_{\bar{m}}{h}_{\bar{j}\hat{j}}(t){h}_{\bar{j}j}(t+T_{s})+\frac{S_{n}-S_{m}}{2}\sum_{i=1}^{N}{h}^{2}_{ij}(t)}{\sqrt{\sum_{i=1}^{N}{h}^{2}_{ij}(t)\sigma^{2}_{ni,\sss \text{ SM}}(t)}}\right), & \text{for $S_{m}<S_{n}$}.
	\end{cases}
	\end{align}
	\rule{\textwidth}{0.3mm}
\end{figure*}

Consequently, when both the SSK and the CSK symbols are considered, the error probability of the SM-MC system conditioned on that $S^{j}_{m}$ and $S^{\bar{j}}_{\bar{m}}$ are transmitted as the current and previous symbol, is given by 
\begin{align}
P_{e,\sss \text{ SM}}(S^{j}_{m}|S^{\bar{j}}_{\bar{m}})=&1\hspace{-0.1cm}-\hspace{-0.1cm}\Big(\hspace{-0.1cm}1-P_{e,\sss \text{ SSK}}(S^{j}_{m}|S^{\bar{j}}_{\bar{m}})\hspace{-0.1cm}\Big)\Big(1-P_{e,\sss \text{ CSK}}(S^{j}_{m}|S^{\bar{j}}_{\bar{m}})\hspace{-0.1cm}\Big)\nonumber\\
\leqslant&P_{e,\sss \text{ SSK}}(S^{j}_{m}|S^{\bar{j}}_{\bar{m}})+\sum_{m\neq n}\text{Pr}[S_{m}\rightarrow S_{n}|S^{\bar{j}}_{\bar{m}}]\nonumber\\
&-P_{e,\sss \text{ SSK}}(S^{j}_{m}|S^{\bar{j}}_{\bar{m}})\sum_{m\neq n}\text{Pr}[S_{m}\rightarrow S_{n}|S^{\bar{j}}_{\bar{m}}].
\end{align}
Furthermore, if we ignore the negative term in (37), we have
\begin{align}
P_{e,\sss \text{ SM}}(S^{j}_{m}|S^{\bar{j}}_{\bar{m}})
\leqslant&P_{e,\sss \text{ SSK}}(S^{j}_{m}|S^{\bar{j}}_{\bar{m}})+\sum_{m\neq n}\text{Pr}[S_{m}\rightarrow S_{n}|S^{\bar{j}}_{\bar{m}}].
\end{align}

When comparing (37) and (38), we can see that when the signal-to-noise ratio (SNR) is sufficiently high, making the detection of both the SSK and CSK symbols sufficiently reliable, then we can ignore the production term in (37), and directly use (38) to obtain the approximate conditional error probability, as shown in Section IV.

Finally, the average SER of the SM-MC systems can be approximately evaluated as
\begin{align}
P_{e,\sss \text{ SM}}=\frac{1}{M^{2}N^{2}}\sum_{j=1}^{N}\sum_{m=0}^{M-1}\sum_{\bar{j}=1}^{N}\sum_{\bar{m}=0}^{M-1}P_{e,\sss \text{ SM}}(S^{j}_{m}|S^{\bar{j}}_{\bar{m}}).
\end{align}

\begin{table}[b]
	\caption{System parameters}  
	\begin{tabular*}{9cm}{lccc}\toprule 
		Parameter & Variable & Value &Unit   \\ 
		\toprule 
		Diffusion Coefficient &$D$   & $2.2\times10^{-9}$&$m^{2}/s$  \\  
		Link Distance &$d$  & $20$ &$\rm{\mu m}$ \\
		Receiver Radius &$\rho$  & $0.1$&$\rm{\mu m}$  \\ 
		SNR Range &  & [$0$, $20$]&$\rm{dB}$\\   
		SNR Interval &  & $2$&$\rm{dB}$\\ 
		Order of Concentration Symbol& $M$& $[2,4]$\\  
		Order of Space Symbol& $N$& $[2,4]$\\  
		Symbol Sequence Length &  & $10^{6}$\\   
		Replication Times&  & $20$ \\ 
		\bottomrule
	\end{tabular*}  
\end{table}
\section{Numerical Results}
In this section, we present the analytical and simulation results for the SERs of the SM-MC and SSK-MC systems. Performance comparison among various modulation schemes is also provided.

The definition of SNR in SISO-MC has been proposed in \cite{LLYang.MCanalysis.16} when OOK modulation schemes are considered. Here, we generalize the SNR definition to the $N\times N$ SM-MC and MIMO-MC with $M$-ary CSK modulation. Specifically, in SM-MC, the SNR can be defined similar to SISO-MC, as the ratio between the average received power of the desired link from a single transmitted impulse of molecules and the noise power, expressed as
\begin{align}
{\text{SNR}}&=\frac{P_{s}}{P_{n}}=\frac{1}{MN}\sum_{j=1}^{N}\sum_{m=0}^{M-1}S_{m}h_{jj}(t_{p})V_{\sss\rm{RX}}\nonumber\\
&=\bigg(\frac{3}{2\pi e}\bigg)^{\frac{3}{2}}\frac{V_{\sss\rm{RX}}}{Md^{3}}\sum_{m=0}^{M-1}S_{m},
\end{align}
where $S_{m}>0$ should be satisfied. Eq. (40) implies that SNR is merely dependent on the molecule number of molecules released in average by a chemical impulse, the transceiver distance $d$ and the volume of receiver $V_{\sss {\text{RX}}}$. Similarly, the SNR of each link in the $N\times N$ MIMO-MC with $M$-ary CSK can be defined as
\begin{align}
{\text{SNR}}=\bigg(\frac{3}{2\pi e}\bigg)^{\frac{3}{2}}\frac{V_{\sss {\text{RX}}}}{NMd^{3}}\sum_{m=0}^{M-1}S_{m}.
\end{align}
Note that (41) is $1/N$ of the (40), indicating that all the transmitters are activated to emitted chemical pulses in each symbol duration, when $S_{m}>0$. Specifically, when the OOK modulation is used in the $N\times N$ MIMO-MC, only $N/2$ transmitters are activated in average to emitted chemical pulses with $S_{1}$ molecules during each symbol period. Hence, the SNR of each link in the $N\times N$ MIMO-MC is then defined as
\begin{align}
{\text{SNR}}=\bigg(\frac{3}{2\pi e}\bigg)^{\frac{3}{2}}\frac{2V_{\sss {\text{RX}}}}{Nd^{3}}S_{1},
\end{align}
where $S_{1}$ is the number of molecules for transmitting bit ``1''.

=When the transceiver distance $d$ is fixed, the same SNR implies the same molecular energy consumption. We can make relatively fair SER comparison between different modulation schemes of MC with the same transmission rate, under the same energy consumption. In order to achieve this, in our simulation for SM-MC with BCSK, we set $S_{1}=2S_{0}$. By contrast, for SISO-MC with QCSK, we have $S_{0}=0$ and $S_{3}=\frac{3}{2}S_{2}=3S_{1}$. The other system parameters are presented in Table~II. Note that in this section, the SSK-MC with the $2\times2$ and $4\times4$ MIMO architectures, which for brevity are referred to as the BSSK-MC and QSSK-MC.
\begin{figure}[tb]
	\centering
	\includegraphics[width=3.5in]{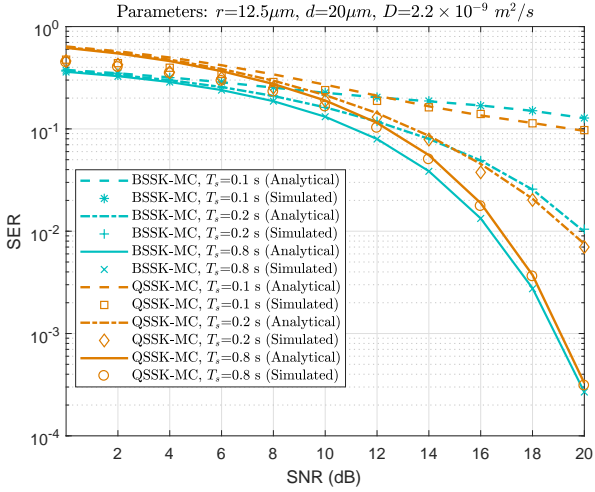}
	\caption{SER comparison between BSSK-MC and QSSK-MC, when receiver separation distance $r=12.5\ \rm{\mu m}$ and different symbol durations $T_{s}=0.1\ \rm{s},0.2\ \rm{s},0.8\ \rm{s}$ are assumed.}
	\label{fig_4}
\end{figure}

Fig. 4 shows the analytical and simulation results for the SER of the BSSK-MC and QSSK-MC, when different symbol durations are considered. Clearly, the analytical SER agrees well with the corresponding simulated SER. In the low SNR region, the SER of QSSK-MC is much higher than that of the BSSK-MC, while in the high SNR region, the SER performance of QSSK-MC is becomes better than that of the BSSK-MC, when $T_{s}=0.1\ \rm{s}$ and $T_{s}=0.2\ \rm{s}$. By contrast, when $T_{s}=0.8\ \rm{s}$, the SER of BSSK-MC is always lower than that of QSSK-MC in the whole SNR region considered. However, the gap between their SER performance decreases with the increase of SNR. As shown in Fig. 4, the SER performance of BSSK-MC and QSSK-MC is highly dependent on the symbol duration $T_{s}$ . When $T_{s}$ increases, the SER reduces due to the fact that the ISI reduces, as $T_{s}$ increases.
\begin{figure}[tb]
	\centering
	\includegraphics[width=3.5in]{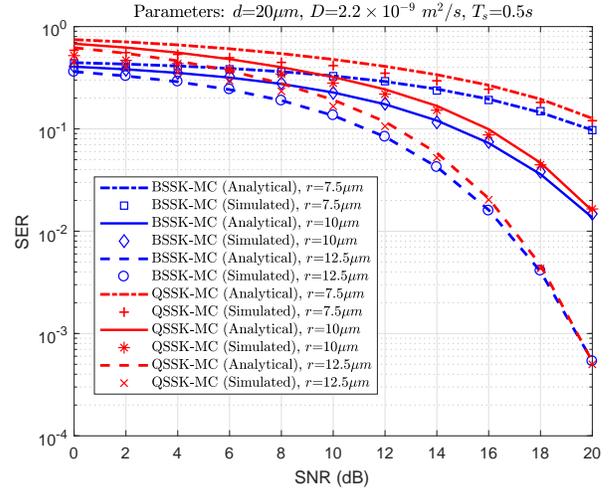}
	\caption{SER comparison between BSSK-MC and QSSK-MC, when $T_{s}=0.5\ \rm{s}$, and different receiver separation distances are considered.}
	\label{fig_5}
\end{figure}

Fig. 5 manifests the analytical and simulation SER of BSSK-MC and QSSK-MC, when different transceiver separation distances are considered. Again, the simulation and analytical results match well, which hence validate our theoretical analysis. As shown in the figure, for a given transceiver separation distance, there is a gap between the SER performance of BSSK-MC and that of QSSK-MC in the low SNR region, with BSSK-MC always outperforming the QSSK-MC. However, as the SNR increases, the SER curves of BSSK-MC and QSSK-MC converge and become nearly the same at high SNR. Therefore, when sufficient source of information molecules, i.e., SNR, is available, higher throughput can be attained by utilizing higher order modulation schemes in the SSK-MC, while achieving the required error performance.
\begin{figure}[tb]
	\centering
	\includegraphics[width=3.5in]{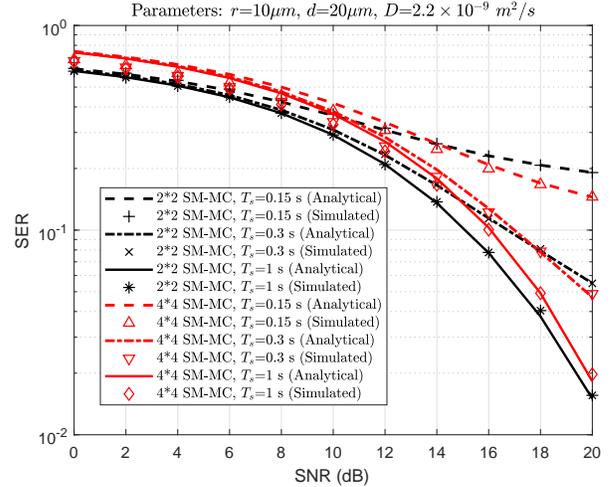}
	\caption{SER comparison between the $2\times2$ SM-MC and $4\times4$ SM-MC, when receiver separation distance $r=10\ \rm{\mu m}$ and different symbol durations $T_{s}=0.15\ \rm{s},0.3\ \rm{s},1\ \rm{s}$ are assumed.}
	\label{fig_6}
\end{figure}

Fig. 6 depicts the theoretical SER upper bound and the simulated SER of the SM-MC, when the $2\times2$ and the $4\times4$ MIMO architectures are respectively considered. In this figure, the transceiver separation distance is fixed to $r=10\ \rm{\mu m}$, whilst the symbol durations $T_{s}$ is set to $0.15\ \rm{s}$, $0.3\ \rm{s}$ or $1\ \rm{s}$. The results show that the upper bound is tight, which becomes tighter as the SNR increases. As shown in Fig. 6, in the low SNR regime, the SER of the $4\times4$ SM-MC is much higher than that of the $2\times2$ SM-MC. By contrast, in the high SNR regime, the SER of the $4\times4$ SM-MC is better than that of the $2\times2$ SM-MC, when $T_{s}=0.15\ \rm{s}$ or $T_{s}=0.3\ \rm{s}$. Furthermore, when $T_{s}=1\ \rm{s}$, the SER of the BSSK-MC is always lower than that of the QSSK-MC in the SNR regime considered. The explanation for these results are similar to that for the results shown in Fig. 4.
\begin{figure}[tb]
	\centering
	\includegraphics[width=3.5in]{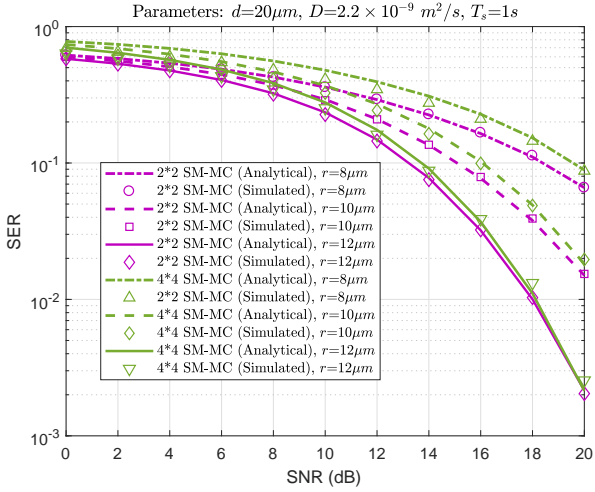}
	\caption{SER comparison between the $2\times2$ SM-MC and $4\times4$ SM-MC, when $T_{s}=1\ \rm{s}$, and $r=8\ \rm{\mu m},10\ \rm{\mu m},12\ \rm{\mu m}$ are assumed.}
	\label{fig_7}
\end{figure}

Fig. 7 also demonstrates the theoretical SER upper bound and the simulation SER results of the SM-MC, where the symbol duration is set to $T_{s}=1\ \rm{s}$, while various transceiver separation distances are considered. Similar to Fig. 6, the SER upper bounds match well with their corresponding results obtained from simulations. There is also an SER gap between the $2\times2$ SM-MC and its $4\times4$ counterpart. However, when the transceiver separation distance increases, the SER difference evaluated in $\rm{dB}$ reduces. We should note that when the link distance $d$ and the symbol duration $T_{s}$ are fixed, the SER performance of SM-MC is mainly determined by the separation distance $r$ between the receivers.
\begin{figure}[tb]
	\centering
	\includegraphics[width=3.5in]{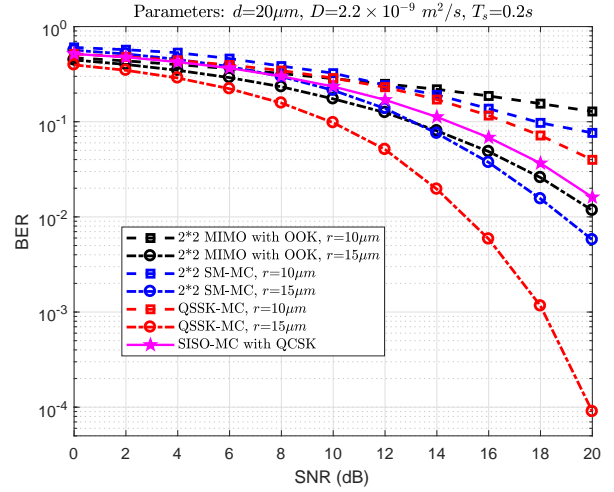}
	\caption{SER comparison of the SM-MC and SSK-MC with that of the MIMO-MC using OOK modulation and SISO-MC using QCSK modulation, when $T_{s}=0.2\ \rm{s}$, and $r=10\ \rm{\mu m}$ or $15\ \rm{\mu m}$.}
	\label{fig_8}
\end{figure}
\begin{figure}[tb]
	\centering
	\includegraphics[width=3.5in]{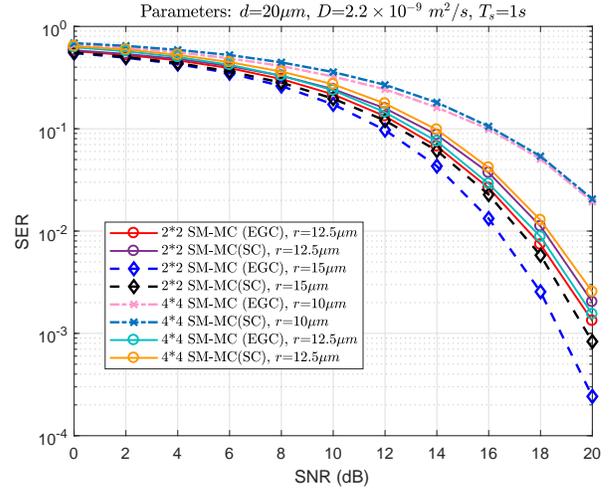}
	\caption{SER performance of the SM-MC with EGC-assisted and SC-assisted detection schemes, when $T_{s}=1\ \rm{s}$, and $r=10\ \rm{\mu m},12.5\ \rm{\mu m}$ or $15\ \rm{\mu m}$.}
	\label{fig_9}
\end{figure}

To verify the effectiveness of our proposed SSK-MC and SM-MC, we compare the simulated SER performance of the SISO-MC with QCSK, $2 \times 2$ MIMO-MC with OOK, $4 \times 4$ SSK-MC and the $2 \times 2$ SM-MC in Fig. 8. All of them have the same transmission rate of 2 bits per symbol. As shown in Fig. 8, when $r=15\ \rm{\mu m}$, the QSSK-MC outperforms all the other schemes and achieves the best SER performance. Both Both the SSK-MC and SM-MC benefit from the employment of space modulation, making them significantly surpass the $2 \times 2$ MIMO-MC in the high SNR regime. The SISO-MC with QCSK attains the worst SER performance among the schemes considered. However, the situation is completely different when the separation distance is reduced to $r=10\ \rm{\mu m}$. In this case, the SISO-MC with QCSK achieves the lowest SER, and outperforms all the other MIMO schemes. As shown in Fig. 8, the $2 \times 2$ MIMO-MC with OOK has similar SER performance as the QSSK-MC and the $2 \times 2$ SM-MC when the SNR is between $0\ \rm{dB}$ and $12\ \rm{dB}$. When for the increasing SNR, its SER performance becomes poorer than that of the other schemes.

The results in Fig. 8 implies that the separation distance $r$ is important for the achievable SER performance, especially when comparing the modulation schemes in SISO-MC. When $r$ is appropriately selected, the MIMO-MC schemes are capable of outperforming SISO-MC with QCSK, owning to the significant ILI reduction in the MIMO-MC schemes, which is beneficial to transmitting information in the space domain. By contrast, if $r$ is not sufficiently large, the advantage of using space symbol vanishes due to the high ILI.

In Fig. 9, we compare the SER performance of the SC-assisted SM-MC of (28) with that of the EGC-assisted SM-MC of (29). From Fig. 9, we observe that the SER performance of EGC-assisted SM-MC is better than that of SC-assisted SM-MC, when there is only a single receiver utilized to decode the concentration symbol. For the $4 \times 4$ SM-MC with $r=10\ \rm{\mu m}$, EGC-assisted SM-MC slightly outperforms SC-assisted SM-MC, when the SNR is lower than about $16 \ \rm{dB}$, while their performance converges at high SNR. When $r=12.5\ \rm{\mu m}$, and the $2 \times 2$ and $4 \times 4$ SM-MC are considered, we observe that the EGC-assisted SM-MC always outperforms its SC-assisted counterpart. Additionally, when the same combining scheme is employed at the receivers, a small-scale SM-MC with fewer links outperforms a large-scale SM-MC with more links. Fig.9 also shows, when $2 \times 2$ SM-MC with $r=15\ \rm{\mu m}$ is considered, the SER performance of EGC-assisted SM-MC becomes increasingly better than that of SC-assisted SM-MC, and approximately $1.5\ \rm{dB}$ gain can be obtained at the SER of $10^{-3}$.

Based on the aforementioned results, in order to improve the SER performance of the SSK-MC and that of SM-MC, the following approaches may be applied. Firstly, the symbol duration $T_{s}$ may be increased to reduce ISI, which hence improves the SER of the SSK-MC and SM-MC systems, but at the cost of the transmission rate. Secondly, the separation distance between adjacent transmitters and that between adjacent receivers may be increased for mitigating the ILI. This is an effective method to improve the SER performance provided that tehre are spaces for transmitter and receiver deployment. Furthermore, the number of molecules per pulse maybe increased, which increases SNR, and therefore enhances the reliability of the information transmission in SSK-MC and the SM-MC system, but at the cost of energy consumption.

\section{Conclusions and Future Work}
In this paper, we have proposed a SM-MC scheme and its special case of SSK-MC by introducing new degrees of freedom from the spatial domain to transmit more information in MC. Our studies showed that when the separation distance between transmit/receive nanomachines is appropriate, additional information conveyed in the space domain is achievable, and our proposed SM-MC and SSK-MC schemes are capable of outperforming the conventional SISO-MC and MIMO-MC scheme. The reason behind is that our proposed SM-MC can effectively eliminate the ILI, which usually severely affects the performance of the conventional MIMO-MC. Moreover, it is an energy-efficient system where only one nanomachine transmitter is activated during each symbol period, befitting the twofold goal in small scale communication systems of low energy consumption and low-complexity.

In this paper, we have also presented theoretical analysis and some simulation results under the assumption of the homogeneous propagation medium as done in the literature. This may not accord with a realistic situation where various medium exist in the communication channel. In our future work, we will consider the diffusion based MC with more realistic channel modelling, as well as the design of corresponding MIMO-MC transceivers.

\begin{appendices}
\section{  }
Let us denote $\text{Pr}_{\sss \text{ SSK}}[i|S^{\bar{j}}_{\bar{m}}]$ as the probability that the $i$-th receiver senses the maximum molecular concentration, which can be expressed as
	\begin{align}
	\text{Pr}_{\sss \text{ SSK}}[i|S^{\bar{j}}_{\bar{m}}]=\text{Pr}\Big[\hspace{-0.05cm}\big(y_{i,\sss \text{ SM}}(t)\hspace{-0.1cm}>\hspace{-0.05cm}y_{1,\sss \text{ SM}}(t)\hspace{-0.05cm}\big)\hspace{-0.1cm}\cap\hspace{-0.1cm}\cdots\hspace{-0.1cm}\cap\hspace{-0.1cm}\big(y_{i,\sss \text{ SM}}(t)\hspace{-0.1cm}>\hspace{-0.05cm}y_{N,\sss \text{ SM}}(t)\hspace{-0.05cm}\big)\hspace{-0.05cm}\Big],
	\end{align}
when given that the previously sent molecular symbol is $S^{\bar{j}}_{\bar{m}}$. In order to derive the error detection probability of the space symbol, let us first derive the correct detection probability of the space symbol, when $S_{m}$ molecules are released from the $j$-th transmitter. Based on (43), this probability can be expressed as
\begin{align}
P_{c,\sss \text{ SSK}}(S^{j}_{m}|S^{\bar{j}}_{\bar{m}})&=\text{Pr}_{\sss \text{ SSK}}[j|S^{\bar{j}}_{\bar{m}}]\nonumber\\
&=\text{Pr}\hspace{-0.05cm}\Big[\hspace{-0.075cm}\big(y_{j,\sss \text{ SM}}(t)\hspace{-0.1cm}>\hspace{-0.1cm}y_{1,\sss \text{ SM}}(t)\hspace{-0.075cm}\big)\hspace{-0.1cm}\cap\hspace{-0.1cm}\cdots\hspace{-0.1cm}\cap\hspace{-0.1cm}\big(y_{j,\sss \text{ SM}}(t)\hspace{-0.1cm}>\hspace{-0.1cm}y_{N,\sss \text{ SM}}(t)\hspace{-0.075cm}\big)\hspace{-0.075cm}\Big],
\end{align}
which is the probability that the $j$-th receiver senses the maximum concentration. When the reception processes of the receiver nanomachines are assumed to be independent, (44) can be expressed as
\begin{align}
P_{c,\sss \text{ SSK}}(S^{j}_{m}|S^{\bar{j}}_{\bar{m}})=\prod_{i\neq j}\text{Pr}[y_{j,\sss \text{ SM}}(t)-y_{i,\sss \text{ SM}}(t)>0|S^{\bar{j}}_{\bar{m}}].
\end{align}

Furthermore, when given $S^{\bar{j}}_{\bar{m}}$, $y_{j,\sss \text{ SM}}(t)-y_{i,\sss \text{ SM}}(t)$ can be written as
\begin{align}
y_{ji|S^{\bar{j}}_{\bar{m}}}=&S_{m}\big(h_{jj}(t)-h_{ij}(t)\big)+I_{j,\sss \text{ SM}}(t)-I_{i,\sss \text{ SM}}(t)\nonumber\\&+n_{j,\sss \text{ SM}}(t)-n_{i,\sss \text{ SM}}(t).
\end{align}
According to (25), $y_{ji|S^{\bar{j}}_{\bar{m}}}$ follows the Normal distribution of
\begin{align}
y_{ji|S^{\bar{j}}_{\bar{m}}}\sim \mathcal{N} \left(\mu_{ji|S^{\bar{j}}_{\bar{m}}},\sigma^{2}_{ji|S^{\bar{j}}_{\bar{m}}}\right).
\end{align}
When given $i\neq j$, and according to \cite{Pat.NormalDist.96}, $\mu_{ji|S^{\bar{j}}_{\bar{m}}}$ and $\sigma^{2}_{ji|S^{\bar{j}}_{\bar{m}}}\hspace{-0.15cm}$ are given by
\begin{align}
\mu_{ji|S^{\bar{j}}_{\bar{m}}}\hspace{-0.15cm}=\hspace{-0.15cm}
\begin{cases}
\vspace{0.15cm} 
\
\hspace{-0.225cm}S_{m}\hspace{-0.075cm}\big(h_{jj}\hspace{-0.05cm}(t)\hspace{-0.1cm}-\hspace{-0.1cm}h_{ij}\hspace{-0.05cm}(t)\hspace{-0.06cm}\big)\hspace{-0.1cm}-\hspace{-0.1cm}S_{\bar{m}}h_{ii}\hspace{-0.05cm}(t\hspace{-0.1cm}+\hspace{-0.1cm}T_{s}\hspace{-0.05cm}),\hspace{-0.3cm}&\text{for $i= \bar{j}$ and $j\neq \bar{j}$,}\\
\
\vspace{0.15cm}  \hspace{-0.225cm}S_{m}\hspace{-0.075cm}\big(h_{jj}\hspace{-0.05cm}(t)\hspace{-0.1cm}-\hspace{-0.1cm}h_{ij}\hspace{-0.05cm}(t)\hspace{-0.06cm}\big)\hspace{-0.1cm}+\hspace{-0.1cm}S_{\bar{m}}h_{jj}\hspace{-0.05cm}(t\hspace{-0.1cm}+\hspace{-0.1cm}T_{s}\hspace{-0.05cm}),\hspace{-0.3cm}&\text{for $i\neq \bar{j}$ and $j= \bar{j}$,}\\
\ \hspace{-0.225cm}S_{m}\hspace{-0.075cm}\big(h_{jj}\hspace{-0.05cm}(t)\hspace{-0.1cm}-\hspace{-0.1cm}h_{ij}\hspace{-0.05cm}(t)\hspace{-0.06cm}\big), \hspace{-0.3cm}&\text{for $i\neq \bar{j}$ and $j\neq \bar{j}$,}
\end{cases}
\end{align}
\begin{align}
\sigma^{2}_{ji|S^{\bar{j}}_{\bar{m}}}\hspace{-0.15cm}=\hspace{-0.15cm}
\begin{cases}
\vspace{0.15cm}
\ \hspace{-0.25cm}\frac{S_{m}\big(h_{jj}(t)+h_{ij}(t)\big)+S_{\bar{m}}h_{ii}(t+T_{s})}{V_{\sss\text{RX}}},&\text{for $i= \bar{j}$ and $j\neq \bar{j}$,}\\
\vspace{0.15cm}
\ \hspace{-0.25cm}\frac{S_{m}\big(h_{jj}(t)+h_{ij}(t)\big)+S_{\bar{m}}h_{jj}(t+T_{s})}{V_{\sss\text{RX}}},&\text{for $i\neq \bar{j}$ and $j= \bar{j}$,}\\
\ \hspace{-0.25cm}\frac{S_{m}\big(h_{jj}(t)+h_{ij}(t)\big)}{V_{\sss\text{RX}}},&\text{for $i\neq \bar{j}$ and $j\neq \bar{j}$,}
\end{cases}
\end{align}
respectively. Therefore, we have
\begin{align}
\text{Pr}[y_{ji|S^{\bar{j}}_{\bar{m}}}\hspace{-0.1cm}>\hspace{-0.1cm}0]\hspace{-0.1cm}&=\hspace{-0.2cm}\int_{0}^{+\infty}\hspace{-0.5cm}\frac{1}{\sqrt{2\pi}\sigma_{ji|S^{\bar{j}}_{\bar{m}}}}\exp\hspace{-0.1cm}\left(\hspace{-0.15cm}-\frac{(y_{ji|S^{\bar{j}}_{\bar{m}}}-\mu_{ji|S^{\bar{j}}_{\bar{m}}})^{2}}{2\sigma^{2}_{ji|S^{\bar{j}}_{\bar{m}}}}\hspace{-0.125cm}\right)\hspace{-0.125cm}dy_{ji|S^{\bar{j}}_{\bar{m}}}\nonumber\\
&=Q\left(-\frac{\mu_{ji|S^{\bar{j}}_{\bar{m}}}}{\sigma_{ji|S^{\bar{j}}_{\bar{m}}}}\right).
\end{align}
Finally, when substituting (50) into (45), the probability of correct detection of the space symbol $j$ can be formulated as
\begin{align}
P_{c,\sss \text{ SSK}}(S^{j}_{m}|S^{\bar{j}}_{\bar{m}})=\prod_{i\neq j}Q\left(-\frac{\mu_{ji|S^{\bar{j}}_{\bar{m}}}}{\sigma_{ji|S^{\bar{j}}_{\bar{m}}}}\right).
\end{align}

\end{appendices}
\bibliographystyle{IEEEtran} 
\bibliography{IEEEabrv,SMMCREF}

\begin{thebibliography}{10}
\providecommand{\url}[1]{#1}
\csname url@samestyle\endcsname
\providecommand{\newblock}{\relax}
\providecommand{\bibinfo}[2]{#2}
\providecommand{\BIBentrySTDinterwordspacing}{\spaceskip=0pt\relax}
\providecommand{\BIBentryALTinterwordstretchfactor}{4}
\providecommand{\BIBentryALTinterwordspacing}{\spaceskip=\fontdimen2\font plus
\BIBentryALTinterwordstretchfactor\fontdimen3\font minus
  \fontdimen4\font\relax}
\providecommand{\BIBforeignlanguage}[2]{{%
\expandafter\ifx\csname l@#1\endcsname\relax
\typeout{** WARNING: IEEEtran.bst: No hyphenation pattern has been}%
\typeout{** loaded for the language `#1'. Using the pattern for}%
\typeout{** the default language instead.}%
\else
\language=\csname l@#1\endcsname
\fi
#2}}
\providecommand{\BIBdecl}{\relax}
\BIBdecl

\bibitem{Shorey.Pheromone.13}
H.~H. Shorey, \emph{Animal Communication by Pheromones}.\hskip 1em plus 0.5em
  minus 0.4em\relax Academic Press, 2013.

\bibitem{Akyildiz13}
I.~Akyildiz, F.~Fekri, R.~Sivakumar, C.~Forest, and B.~Hammer, ``{MONACO}:
  {Fundamentals} of molecular nano-communication networks,'' \emph{IEEE
  Wireless Commun.}, vol.~19, no.~5, pp. 12--18, Oct. 2012.

\bibitem{Han.Signalling.17}
J.~T. Hancock, \emph{Cell Signalling}.\hskip 1em plus 0.5em minus 0.4em\relax
  Oxford University Press, 2017.

\bibitem{Tadashi.MC.05}
T.~Nakano \emph{et~al.}, ``Molecular communication for nanomachines using
  intercellular calcium signaling,'' in \emph{Proc. 5th IEEE Conf. on
  Nanotechnology}, Nagoya, Japan, July 2005, pp. 478--481.

\bibitem{Farsad.MCSurvey.16}
N.~Farsad \emph{et~al.}, ``A comprehensive survey of recent advancements in
  molecular communication,'' \emph{IEEE Commun. Survey \& Tut.}, vol.~18,
  no.~3, pp. 1887--1919, third quarter 2016.

\bibitem{Wang.Nanomachine.13}
J.~Wang, \emph{Nanomachines: Fundamentals and Applications}.\hskip 1em plus
  0.5em minus 0.4em\relax Wiley, 2013.

\bibitem{Tadashi.10year.17}
T.~Nakano, ``Molecular communication: A 10 year retrospective,'' \emph{IEEE
  Trans. Mol. Biol. Multi-Scale Commun.}, vol.~3, no.~2, pp. 71--78, June 2017.

\bibitem{Suzuki.MCmodelling.17}
J.~Suzuki, T.~Nakano, and M.~J. Moore, \emph{Modeling, Methodologies and Tools
  for Molecular and Nano-scale Communications: Modeling, Methodologies and
  Tools}.\hskip 1em plus 0.5em minus 0.4em\relax Springer International
  Publishing, 2017.

\bibitem{Kuran.ModuMC.11}
M.~S. Kuran, H.~B. Yilmaz, T.~Tugcu, and I.~F. Akyildiz, ``Modulation
  techniques for communication via diffusion in nanonetworks,'' in \emph{Proc.
  IEEE Int. Conf. on Commun. (ICC)}, Kyoto, Japan, June 2011, pp. 1--5.

\bibitem{Kim.Isomodu.13}
N.~R. Kim and C.-B. Chae, ``Novel modulation techniques using isomers as
  messenger molecules for nano communication networks via diffusion,''
  \emph{IEEE J. Sel. Areas Commun.}, vol.~31, no.~12, pp. 847--856, Dec. 2013.

\bibitem{Llatser.PPM.13}
I.~Llatser, A.~Cabellos-Aparicio, M.~Pierobon, and E.~Alarcon, ``Detection
  techniques for diffusion-based molecular communication,'' \emph{IEEE J. Sel.
  Areas Commun.}, vol.~31, no.~12, pp. 726--734, Dec. 2013.

\bibitem{Arjmandi.MCSK.13}
H.~Arjmandi, A.~Gohari, M.~N. Kenari, and F.~Bateni, ``Diffusion-based
  nanonetworking: A new modulation technique and performance analysis,''
  \emph{IEEE Commun. Lett.}, vol.~17, no.~4, pp. 645--648, Apr. 2013.

\bibitem{Kabir.DMoSK.15}
M.~H. Kabir, S.~M.~R. Islam, and K.~S. Kwak, ``{D-MoSK} modulation in molecular
  communications,'' \emph{IEEE Trans. Nanobiosci.}, vol.~14, no.~6, pp.
  680--683, Sept. 2015.

\bibitem{Mosayebi.typesignmodu.18}
R.~Mosayebi, A.~Gohari, M.~Mirmohseni, and M.~Nasiri-Kenari, ``Type-based sign
  modulation and its application for {ISI} mitigation in molecular
  communication,'' \emph{IEEE Trans. on Commun.}, vol.~66, no.~1, pp. 180--193,
  Jan. 2018.

\bibitem{KC.frontierWC.12}
P.~C. Yeh \emph{et~al.}, ``A new frontier of wireless communication theory:
  {Diffusion-based} molecular communications,'' \emph{IEEE Wirel. Commun.},
  vol.~19, no.~5, pp. 28--35, Oct. 2012.

\bibitem{KC.MIMOMC.12}
L.~S. Meng, P.~C. Yeh, K.~C. Chen, and I.~F. Akyildiz, ``{MIMO} communications
  based on molecular diffusion,'' in \emph{Proc. IEEE Global Commun. Conf.
  (GLOBECOM)}, Anaheim, CA, USA, Dec. 2012, pp. 5380--5385.

\bibitem{Tadashi.MCbook.13}
T.~Nakano, A.~W. Eckford, and T.~Haraguchi, \emph{Molecular
  Communication}.\hskip 1em plus 0.5em minus 0.4em\relax Cambridge University
  Press, 2013.

\bibitem{Farsad.Prototype.13}
N.~Farsad, W.~Guo, and A.~W. Eckford, ``Tabletop molecular communication: Text
  messages through chemical signals,'' \emph{PLOS ONE}, vol.~8, pp. 1--13, Dec.
  2013.

\bibitem{Chae.MCMIMOproto.16}
B.~H. Koo \emph{et~al.}, ``Molecular {MIMO}: From theory to prototype,''
  \emph{IEEE J. Sel. Areas Commun.}, vol.~34, no.~3, pp. 600--614, Mar. 2016.

\bibitem{Rou.CEMIMOMC.17}
S.~M. Rouzegar and U.~Spagnolini, ``Channel estimation for diffusive {MIMO}
  molecular communications,'' in \emph{Proc. European Conf. on Networks and
  Communi. (EuCNC)}, Oulu, Finland, June 2017, pp. 1--5.

\bibitem{Damrath.Spatialcoding.17}
M.~Damrath, H.~B. Yilmaz, C.-B. Chae, and P.~A. Hoeher, ``Spatial coding
  techniques for molecular {MIMO},'' in \emph{Proc. IEEE Information Theory
  Workshop (ITW)}, Kaohsiung, Taiwan, Nov. 2017, pp. 324--328.

\bibitem{Lee.MLMIMOMC.17}
C.~Lee, H.~B. Yilmaz, C.-B. Chae, N.~Farsad, and A.~Goldsmith, ``Machine
  learning based channel modeling for molecular {MIMO} communications,'' in
  \emph{Proc. IEEE 18th International Workshop on Signal Processing Advances in
  Wireless Communi. (SPAWC)}, Sapporo, Japan, July 2017, pp. 1--5.

\bibitem{WG.MCSIMOsyn.18}
Z.~Luo \emph{et~al.}, ``One symbol blind synchronization in {SIMO} molecular
  communication systems,'' \emph{IEEE Wireless Commun. Lett.}, vol.~PP, no.~99,
  pp. 1--1, 2018.

\bibitem{Mesleh.SM.08}
R.~Y. Mesleh \emph{et~al.}, ``Spatial modulation,'' \emph{IEEE Trans. Veh.
  Technol.}, vol.~57, no.~4, pp. 2228--2241, July 2008.

\bibitem{Jeganathan.SSK.09}
J.~Jeganathan, A.~Ghrayeb, L.~Szczecinski, and A.~Ceron, ``Space shift keying
  modulation for {MIMO} channels,'' \emph{IEEE Trans. Wireless Commun.},
  vol.~8, no.~7, pp. 3692--3703, July 2009.

\bibitem{Wen.IMsurvey.17}
E.~Basar \emph{et~al.}, ``Index modulation techniques for next-generation
  wireless networks,'' \emph{IEEE Access}, vol.~5, pp. 16\,693--16\,746, 2017.

\bibitem{Yilmaz.14}
H.~B. Yilmaz, A.~C. Heren, T.~Tugcu, and C.-B. Chae, ``Three-dimensional
  channel characteristics for molecular communications with an absorbing
  receiver,'' \emph{IEEE Commun. Letters}, vol.~18, no.~6, pp. 929--932, June
  2014.

\bibitem{Zamiri.CDMAMC.16}
Y.~Zamiri-Jafarian, S.~Gazor, and H.~Zamiri-Jafarian, ``Molecular code division
  multiple access in nano communication systems,'' in \emph{Proc. IEEE Wireless
  Communi. and Netw. Conf.}, Apr. 2016, pp. 1--6.

\bibitem{Tepekule.Memoryless.15}
B.~Tepekule \emph{et~al.}, ``{ISI} mitigation techniques in molecular
  communication,'' \emph{IEEE Trans. Mol. Biol. Multi-Scale Commun.}, vol.~1,
  no.~2, pp. 202--216, June 2015.

\bibitem{Dk.receiverMC.13}
D.~Kilinc and O.~B. Akan, ``Receiver design for molecular communication,''
  \emph{IEEE J. Sel. Areas Commun.}, vol.~31, no.~12, pp. 705--714, Dec. 2013.

\bibitem{MM.studyofCEMC.14}
M.~U. Mahfuz, D.~Makrakis, and H.~T. Mouftah, ``A comprehensive study of
  sampling-based optimum signal detection in concentration-encoded molecular
  communication,'' \emph{IEEE Trans. Nanobiosci.}, vol.~13, no.~3, pp.
  208--222, Sept. 2014.

\bibitem{ANoel.Passivereceiver.16}
A.~Noel, Y.~Deng, D.~Makrakis, and A.~Hafid, ``Active versus passive: Receiver
  model transforms for diffusive molecular communication,'' in \emph{Proc. IEEE
  Global Communi. Conf. (GLOBECOM)}, Washington, DC, USA, Dec. 2016, pp. 1--6.

\bibitem{GeC.adadetec.18}
G.~Chang, L.~Lin, and H.~Yan, ``Adaptive detection and {ISI} mitigation for
  mobile molecular communication,'' \emph{IEEE Trans. Nanobiosci.}, vol.~17,
  no.~1, pp. 21--35, Mar. 2018.

\bibitem{GD.AmpliDetect.17}
G.~D. Ntouni and G.~K. Karagiannidis, ``Comparison of amplitude detection
  techniques for passive receivers in molecular communications,'' in
  \emph{Proc. 6th International Conf. on Modern Circuits and Systems
  Technologies (MOCAST)}, Thessaloniki, Greece, May 2017, pp. 1--4.

\bibitem{AN.Enzyme.14}
A.~Noel, K.~C. Cheung, and R.~Schober, ``Improving receiver performance of
  diffusive molecular communication with enzymes,'' \emph{IEEE Trans.
  Nanobiosci.}, vol.~13, no.~1, pp. 31--43, Mar. 2014.

\bibitem{MengLS.receiverdesign.14}
L.~S. Meng, P.~C. Yeh, K.~C. Chen, and I.~F. Akyildiz, ``On receiver design for
  diffusion-based molecular communication,'' \emph{IEEE Trans. Signal
  Process.}, vol.~62, no.~22, pp. 6032--6044, Nov. 2014.

\bibitem{LLY.PSM.11}
L.~L. Yang, ``Transmitter preprocessing aided spatial modulation for
  multiple-input multiple-output systems,'' in \emph{Proc. IEEE 73rd Vehicular
  Technology Conf. (VTC Spring)}, May 2011, pp. 1--5.

\bibitem{JJ.optdetectSM.08}
J.~Jeganathan, A.~Ghrayeb, and L.~Szczecinski, ``Spatial modulation: optimal
  detection and performance analysis,'' \emph{IEEE Commun. Lett.}, vol.~12,
  no.~8, pp. 545--547, Aug. 2008.

\bibitem{RZ.GPSM.13}
R.~Zhang, L.~L. Yang, and L.~Hanzo, ``Generalised pre-coding aided spatial
  modulation,'' \emph{IEEE Trans. Wireless Commun.}, vol.~12, no.~11, pp.
  5434--5443, Nov. 2013.

\bibitem{LLYang.MCanalysis.16}
L.~Shi and L.~L. Yang, ``Diffusion-based molecular communications: Inter-symbol
  interference cancellation and system performance,'' in \emph{Proc. IEEE/CIC
  International Conf. on Communications in China (ICCC)}, Chengdu, China, July
  2016, pp. 1--6.

\bibitem{Pat.NormalDist.96}
J.~Patel and C.~Read, \emph{Handbook of the Normal Distribution, Second
  Edition}.\hskip 1em plus 0.5em minus 0.4em\relax Taylor \& Francis, 1996.

\end{thebibliography}
\end{document}